\documentclass[twocolumn,showpacs,preprintnumbers,amsmath,amssymb]{revtex4}
\usepackage{epsfig}

\def\bea{\begin{eqnarray}}
\def\eea{\end{eqnarray}}

\def\d{\delta}
\def\p{\partial} 
\def\nn{\nonumber}

\def\la{\langle}
\def\ra{\rangle}

\def\om{\omega}
\def\Om{\Omega}
\def\n{\eta}
\def\g{\gamma}

\def\nt{\tilde{\eta}}
\def\ft{\tilde{f}}

\def\f{\frac}

\def\s{\sigma}

\def\G{\Gamma}

\def\mT{\mathcal{T}}

\begin{document}

\title{ Two simple models of classical heat pumps}
\author{Rahul Marathe$^1$, Arun Jayannavar$^2$, Abhishek Dhar$^1$  }
\affiliation{$^1$Raman Research Institute, Bangalore  560080, India}
\affiliation{$^2$Institute of Physics, Bhubaneswar 751005,India }
\date{\today} 

\begin{abstract}
Motivated by recent studies on models of  particle and heat quantum 
pumps, we study similar simple classical models  and examine the
possibility of heat pumping. Unlike many of the usual ratchet models of
molecular engines,  the models we study do not have particle transport.
We consider a two-spin system and a coupled oscillator system
which exchange heat with multiple heat reservoirs and which are acted upon by
periodic forces. The simplicity of our models allows accurate
numerical and exact solutions and unambiguous interpretation of
results. We demonstrate that while both our models seem to be built
on similar principles, one is able to function as a heat pump (or
engine) while the other is not.   
\end{abstract}      

\maketitle

The idea of constructing miniature versions of engines, motors and
pumps has been an interesting one. The earliest theoretical construct
of such a device is  probably Feynman's pawl-and-ratchet model
discussed in \cite{FeynmanLect}. In this article  
Feynman uses this simple microscopic model to demonstrate why a
Maxwell's demon cannot work. In the same article he also shows how
this model can be used  to construct a microscopic heat
engine and discusses its efficiency. There have been a number of
recent detailed studies on the pawl-and-ratchet model and some
subtle flaws in Feynman's original arguments have been pointed out
\cite{Parrondo96,Sekimoto97}.  
A different class of ratchet models have also been studied in 
\cite{Magnasco93}.
In these models Brownian particles, kept in an 
asymmetric periodic potential and acted upon by periodic
time-dependent forces, are found to exhibit directed motion. A number
of variations of this model has been studied \cite{FeldmannHeatEngine96}. Among
its applications it has been proposed that this could provide a
mechanism of transport of motors in biological cells \cite{Astumian97}.

Ratchet models which work on somewhat different principles are 
models of quantum pumps which are recently being studied theoretically 
\cite{Buttiker94} and have also been experimentally realized  
\cite{SwitkesSciqpumpexp99}. 
A typical and simple example of such a device would be 
two coupled quantum dots each separately in contact with particle
reservoirs which are at the same chemical potential. One applies  $AC$
gate voltages $v_1=v_0 \cos (\Omega t)$ and $v_2=v_0 \cos (\Omega t +
\phi)$ to the two dots respectively. This leads to a net flow of
particle current  between the two reservoirs whose sign depends on the
phase $\phi$. An essential ingredient seems to
be a periodic variation of at least two parameters such that the area
enclosed in parameter space is non-zero. Since these pumps also work at
zero temperature it appears that noise is not an essential feature,  
which is unlike the case for usual ratchet models. 
Motivated by the quantum particle pump model, Segal and Nitzan have
proposed a model for a heat pump \cite{Nitzan06}. In this model a
molecule with two allowed 
energy levels interacts with two heat reservoirs kept at different
temperatures. The energy levels are modulated in a periodic way. Thus
unlike the particle pump model here  only a single parameter is 
varied. However an asymmetry is incorporated by taking reservoirs
with different spectral properties and different couplings to the
molecule. This seems to lead to the desired  pumping of heat from the
cold to the hot reservoir.

Motivated by the quantum pump model, in this paper,  we examine
classical models which have the same basic design as the quantum 
version. 
We consider two different models: (i) a spin
system consisting of two Ising spins each driven by periodic magnetic
fields with a phase difference 
and connected to two heat reservoirs and (ii) an oscillator system of 
two interacting particles driven by  periodic forces with a phase
difference and connected to two reservoirs.  In both cases we analyze
the possibility of the 
models to work either as pumps or as engines.
Our main result is that the spin system can work both as a pump and as
an engine. On the other hand the oscillator model fails to perform
either function. 

Our first model consists of two Ising spins  driven by time-dependent
magnetic fields $h_{L}(t)$ and $h_{R}(t)$ respectively and each interacting
with separate heat reservoirs. The Hamiltonian of the  
system is given by
\bea
{\cal H}= -J \s_{1}\s_{2}-h_{L}(t)\s_{1}-h_{R}(t)\s_{2}~,
~~~~~~~~\s_{1,2}=\pm 1 
\eea
where $J$ is the interaction energy between the spins. The magnetic
fields have the forms $h_L(t)=h_0 \cos (\Om t)$ and $h_R(t)= h_0 \cos
(\Om t+\phi)$. The interaction of each spin with the heat baths is
modeled by a stochastic dynamics.  
Here we assume that the time-evolution of the spins is given by
Glauber dynamics \cite{Glauber63}, generalized to the case of two heat baths, with
temperatures $T_L$ and $T_R$. Thus the
Glauber spin flip rates $r^L$ and $r^R$, for the two spins respectively,
are given by  
 $r^L_{\s_1\s_2}=r(1-\g_{L}\s_{1}\s_{2})(1-\nu_{L}\s_{1})$
and $r^R_{\s_1\s_2}=r(1-\g_{R}\s_{1}\s_{2})(1-\nu_{R}\s_{2})$ where 
$\g_{L,R} =\tanh({J}/{k_{B}T_{L,R}} ),~\nu_{L,R}
=\tanh({h_{L,R}}/{k_{B}T_{L,R}})$ 
and $r$ is a rate constant. The master equation for evolution of the
spin distribution function $\hat{P}= [P(+,+,t), P(-,+,t),
  P(+,-,t), P(-,-,t)]^T$ is then given by:
\bea
\f{\p {\hat P}}{\p t} = {\mT}{\hat P}~, \label{masteqn} 
\eea
where\bea
\mT=~~~~~~~~~~~~~~~~~~~~~~~~~~~~~~~~~~~~~~~~~~~~~~~~~~~~~~~~~~~~~~~~~~~~~~\nn \\ \left(\begin{array}{cccc}
-r^L_{++}-r^R_{++}& r^L_{-+} & r^R_{+-}& 0 \\
r^L_{++} & -r^L_{-+}-r^R_{-+} & 0&  r^R_{--} \\
r^R_{++} & 0 & -r^L_{+-}-r^R_{+-}& r^L_{--} \\
0 & r^R_{-+} & r^L_{+-}&-r^L_{--}-r^R_{--} 
\end{array}\right)~. \nn
\eea
\begin{figure}
\begin{center}
\includegraphics[width=3in]{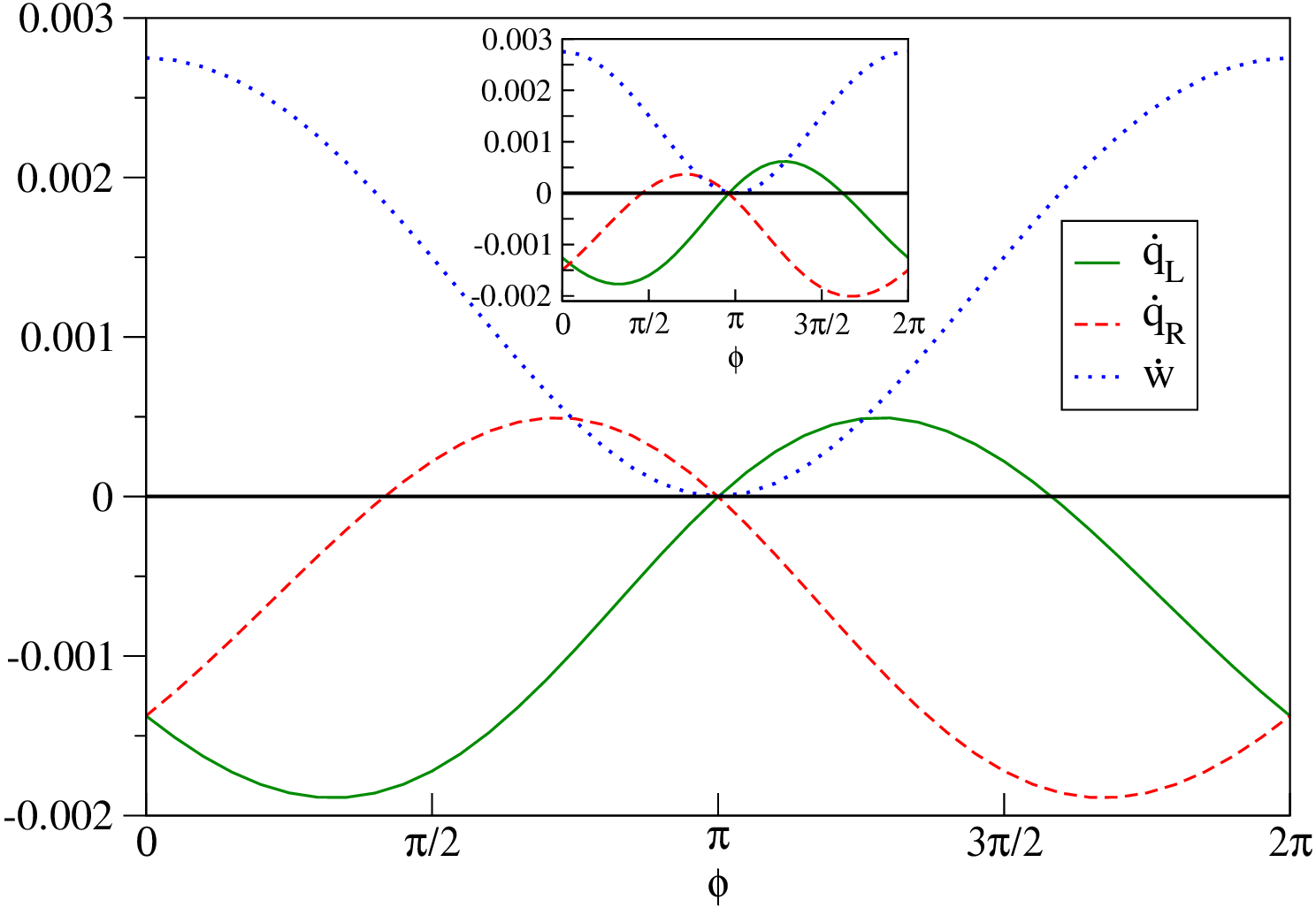}
\caption{(Color online) Plot of $\dot{q}_{L}$, $\dot{q}_{R}$, $\dot{w}$ versus $\phi$
  with both baths at the same temperature. Inset shows the currents
  for the case where the right bath is slightly colder.} 
\label{allref}
\end{center}
\end{figure}
We define $\dot{Q}_{L}$, $\dot{Q}_{R}$ to be the rates (averaged over
the probability ensemble) at which heat is absorbed from the left and
right baths respectively while $\dot{W}_{L}$, $\dot{W}_{R}$ are the
rates at which work is done on the left and right spins by the
external magnetic field. These can be readily expressed in terms of
the spin distribution function and the various transition rates. Thus
we find:
\bea   
\dot{Q}_{L}&=& \sum_{\s_1,\s_2} P(\s_1,\s_2, t) r^L_{\s_1\s_2} \Delta
E_1(\s_1, \s_2) \nn \\
\dot{Q}_{R}&=& \sum_{\s_1,\s_2} P(\s_1,\s_2, t) r^R_{\s_1\s_2} \Delta
E_2(\s_1,\s_2) \nn \\
\dot{W}_{L} &=& - \la \s_1 \ra ~\dot{h}_L= -\dot{h}_L \sum_{\s_1,\s_2}
\s_1 P(\s_1, \s_2, t) \nn \\ 
\dot{W}_{R} &=& - \la \s_2 \ra ~\dot{h}_R= -\dot{h}_R \sum_{\s_1,\s_2}
\s_2 P(\s_1, \s_2, t) ~, \label{curr}
\eea
where $\Delta E_1=2~( J \s_1 \s_2 + h_L \s_1 )$ and   $\Delta E_2=2~(
J \s_1 \s_2 + h_R \s_2 )$   are the energy costs in flipping the first
and second spin respectively. The average energy of the system is
given by $ U = \la \mathcal{H} \ra = \sum_{\s_1, \s_2} \mathcal{H}(\s_1, \s_2) P(\s_1, \s_2,
  t) $. It is easy to verify the  energy conservation
  equation: $\dot{U}=\dot{Q}_L+\dot{Q}_R+\dot{W}_L+\dot{W}_R $. 
Note that even at long times the steady state distribution $\hat{P}$
remains time-dependent. We will also be interested in the
following time averaged rates of heat exchanges and work done,
evaluated in the steady state: $\dot{q}_{L,R}= \f{1}{\tau}\
\int_{0}^{\tau}\dot{Q}_{L,R}~ dt~,
\dot{w}_{L,R}= \f{1}{\tau}\ \int_{0}^{\tau}\dot{W}_{L,R}~ dt$, where
$\tau=2 \pi/\Om$ is the time period of the driving field. 

\begin{figure}
\includegraphics[width=3in]{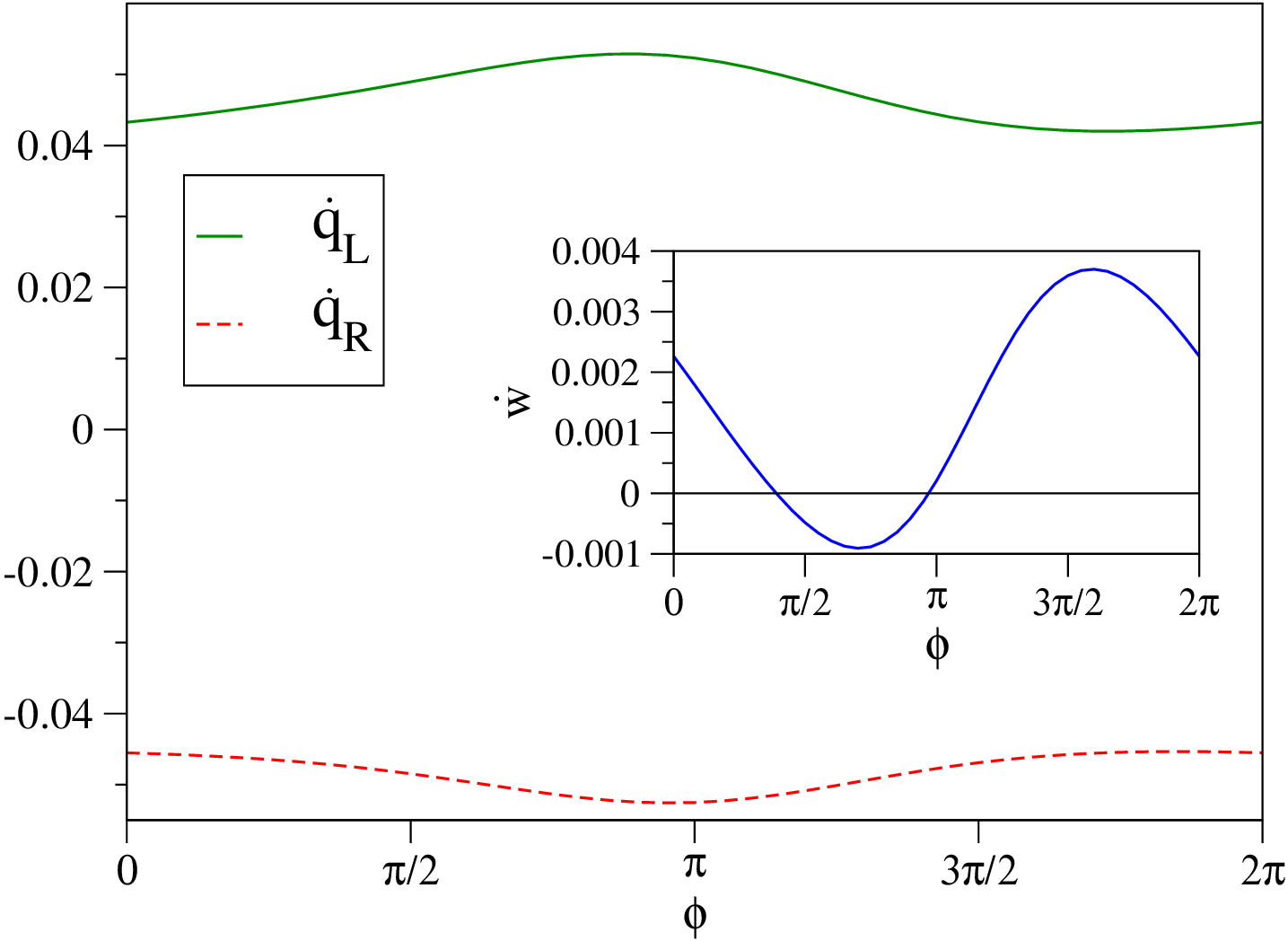}
\caption{(Color online) Plot of $\dot{q}_{L}$, $\dot{q}_{R}$, $\dot{w}$  versus $\phi$
  for  parameter values chosen such that the model performs as an engine.} 
\label{allen}
\end{figure}

We numerically solve the master equation Eq.~(\ref{masteqn}) and then evaluate the various 
steady-state energy exchange rates 
$\dot{q}_{L,R}$ and $\dot{w}_{L,R}$. In all our numerical
calculations we set $r=0.5$ and  $J/k_B=1$ and  all  other quantities are 
measured in these units. In Fig.~(\ref{allref}) we consider the
parameter values $T_L=T_R=0.5$, $h_0=0.25$, $\tau=225$ and plot
$\dot{q}_L, \dot{q}_R$ and 
$\dot{w}=\dot{w}_L+\dot{w}_R$ as functions of the phase $\phi$. 
It can be seen that,  for certain values of the phase, 
both $\dot{q}_L$ and $\dot{q}_R$ are negative while $\dot{w}$ is
positive.  Following our
sign conventions, this means that all the work from the external driving is
getting dissipated into the two baths.  More interestingly 
we find that for certain values of the phase we can get 
$\dot{q}_L >0 $ and  $\dot{q}_R < 0$ which 
means that there is heat flow {\emph{from}}  the left reservoir {\emph{to}} the
right reservoir.  The direction of heat flow can be reversed by
changing the phase. From continuity arguments it is clear that this
model can also sustain heat flow against a small temperature
gradient. Thus the inset of Fig.~(\ref{allref}) shows the currents when the
right reservoir is kept at a slightly lower temperature
$T_R=0.499$. In the absence of any driving we would get a steady
current $\dot{q}_L=-\dot{q}_R=1.41\times 10^{-4}$ from the left to right
reservoir. In the presence of driving and at a phase value $\phi=2.2$ we
get $\dot{q}_R=3.674\times 10^{-4},~\dot{q_L}=-1.025\times 10^{-3}$ which means that heat flows
\emph{out} of the cold  reservoir.  
Thus we see  that our model can perform as a heat pump 
or a refrigerator. Similarly we find that the model can also perform like an
engine and convert heat to work. This can be seen in Fig.~(\ref{allen}) where we
consider the parameter values $T_L=1.0, T_R=0.1$, $h_0=0.25$, $\tau=190$. 
In this case we find that for certain values of $\phi$ we can have 
$\dot{w} < 0$ which means that work is being done on the external force. For typical values of parameters that we have 
tried we find that the efficiency of the engine is quite low. For example for Fig.~(\ref{allen}) with $\phi= 0.7 \pi$, 
we find $\eta = \arrowvert\dot{w} \arrowvert/\dot{q}_{L} = 1.75 \times 10^{-2}$.     

\begin{figure}
\includegraphics[width=3in]{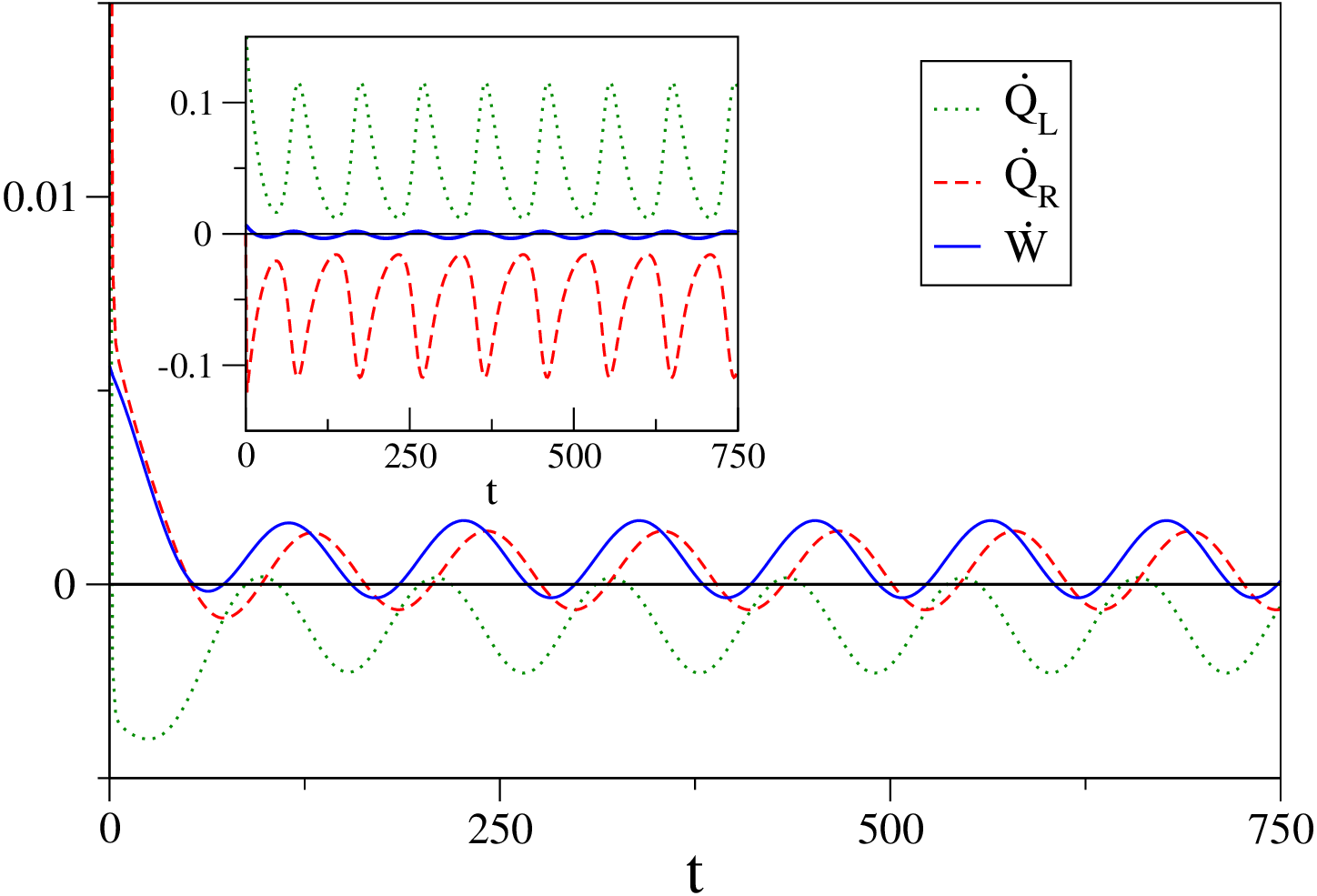}
\caption{(Color online) Plot of  $\dot{Q}_{L}$, $\dot{Q}_{R}$, $\dot{W}$ as a
  function of time for parameters corresponding to pump and engine (inset).}
\label{qlworktref}
\end{figure}

Finally in Fig.(\ref{qlworktref}) we plot the
time-dependent energy transfer rates given by Eq.~(\ref{curr}) for
parameter values corresponding to the refrigerator and engine modes of
operation. In both cases the initial configuration was chosen with
$P(+,+,t=0)=1$. At long times we see that all quantities vary
periodically with time with the same period $\tau$ as the driving
force. Fig.~(\ref{qlworktref}) corresponds to the parameter values
$T_L=0.5,~T_R=0.499,~h_0=0.25,~\tau=225$ and $\phi=2.2$ while the
inset corresponds to the engine parameters 
$T_L=1.0,~ T_R=0.1,~h_0=0.25,~ \tau=190$ and $\phi=2.2$. 

The second model of our engine consists of two particles which
separately interact with two reservoirs kept at different
temperatures. The particles interact with each other and are also
driven by two external periodic forces with a phase difference. 
We consider the system to be described by the Hamiltonian
\bea
{\cal H}&=&\frac{p_1^{2}}{2m}\ +\frac{p_2^{2}}{2m}\ + \frac{1}{2}\ k 
x_1^{2}+\frac{1}{2}\ k x_2^{2}+\frac{1}{2}\ k_c (x_1-x_2)^{2}~\nonumber\\
&-&(f_{L}(t)~x_{1}+f_{R}(t)~x_{2}) \label{oscH}
\eea
The two particles are acted on by external periodic forces given by 
$f_L(t)=f_{0}\cos(\Omega t)$ and $f_R(t)=f_{0}\cos(\Omega t+\phi)$ respectively, 
where $\phi$ is a phase difference. The effect of the 
heat baths at temperatures $T_L$ and $T_R$  
is modeled by Langevin equations. Thus the
equations of motion are 
\bea
m\ddot{x_{1}}&=&-(k+k_{c})x_{1} + k_{c} x_{2} -\gamma \dot{x_{1}} +
\eta_{L} + f_{L}(t)  ~, \nn \\
m\ddot{x_{2}} &=& -(k+k_{c})x_{2} + k_{c} x_{1} -\gamma \dot{x_{2}} +
\eta_{R} + f_{R}(t)  ~, \nn
\eea
where the two noise terms are Gaussian and uncorrelated and satisfy
the  usual fluctuation-dissipation relations $\la \n_{L,R}(t) \n_{L,R} (t')
\ra =2 k_B 
T_{L,R} \g \d (t-t')$.
 Multiplying the two equations above 
by $\dot{x}_1$ and $\dot{x}_2$ respectively and
adding them up we get: $
\dot{\mathcal{H}}=  (-\g \dot{x}_1+\eta_L)\dot{x}_1  +(-\g
\dot{x}_2+\eta_R)\dot{x}_2  
- \dot{f}_{L}(t) x_1 - \dot{f}_{R}(t) x_2$, 
which has the obvious interpretation of an  energy
conservation equation. Averaging over noise we get  
$\dot{U}=\dot{Q}_L+\dot{Q}_R+\dot{W}_L+\dot{W}_R$.  
where the various energy exchange rates have the same interpretations
as in the previous discussion and are given by 
$\dot{Q}_{L} = \la   (-\g \dot{x}_1+\eta_{L})\dot{x}_1 \ra~,
\dot{Q}_{R} = \la  (-\g \dot{x}_2+\eta_{R})\dot{x}_2 \ra~,
\dot{W}_L = -\la \dot{f}_{L} x_1 \ra \nn~,
\dot{W}_R = -\la \dot{f}_{R} x_2 \ra$.
As before we define the average energy transfer rates in the steady state
$\dot{q}_L, \dot{q}_R, \dot{w}_L, \dot{w}_R$. The present model being
linear, it is straightforward to exactly compute these as we now show.

We first obtain the steady-state solutions of the equations of motion. We
write the equations of motion in the following matrix form:
\bea
M \dot{X}=-\Phi X -\G \dot{X} + \eta(t) + f(t) \label{eqmM1mat}
\eea
where $X=[x_1,x_2]^T$, $\eta=[\n_L,\n_R]^T,~f=[f_0\cos(\Om t),f_0
\cos(\Om t+\phi)]^T$,   $M$ and $\Gamma$ are diagonal matrices with
diagonal elements $m$ and $\gamma$ respectively and $\Phi$ is the
force constant matrix. The steady state solution of this equation is:
\bea
X(t)&=&X_N(t)+X_D(t)~, \nn \\
{\rm where~~}X_N(t)&=& \int_{-\infty}^\infty d \om e^{-i \om t} G(\om) \nt
(\om)~, \nn \\
X_D(t)&=& Re[G(\Om) \ft e^{-i \Om t}]~,  \nn \\
{\rm with~~~}G(\om)&=& [\Phi- \om^2 M + i \om \G]^{-1}~, \label{solM2}
\eea
and $\nt=\int_{-\infty}^{\infty} d\om e^{-i \om t}~\eta(t)$, $\ft=\{1, 
e^{-i
  \phi}\}^T$. It is easy to see that the matrix $G(\om)$ has two
independent elements and we denote them as
$A(\om)=G_{11}=G_{22}={[k+k_c-m \om^2 - i \g \om]}/{[(k+k_c-m \om^2-i
    \g \om)^2 -k_c^2]}$ and $ B(\om)=G_{12}=G_{21}={k_c}/{[(k+k_c-m
  \om^2- i \g \om)^2 -k_c^2]}$.
Using the above solution in Eq.~(\ref{solM2}), and after some bit of
algebraic simplifications, we obtain the following results:
\bea
\dot{q}_L&=& -\f{f_0^2 \Om}{2}~[~A_I(\Om)+B_I(\Om) \cos(\phi) +D(\Om) \sin 
(\phi)~] \nn \\ &+& \frac{k_B\g
  k_{c}^{2}({T}_{L}-{T}_{R})}{2(mk_{c}^{2}+(k+k_c)\g^{2})}~,  \nn \\ 
\dot{q}_R&=&  -\f{f_0^2 \Om}{2}~[~A_I(\Om)+B_I(\Om) \cos(\phi) -D(\Om) \sin
  (\phi)~] \nn \\ &+& \frac{k_B\g 
  k_{c}^{2}({T}_{R}-{T}_{L})}{2(mk_{c}^{2}+(k+k_c)\g^{2})}~,  \nn \\
\dot{w}_L &=& \f{f_0^2 \Om}{2}~[~A_I(\Om)+B_I(\Om) \cos (\phi)-B_R(\Om) \sin 
(\phi) ~]~, \nn \\
\dot{w}_R &=& \f{f_0^2 \Om}{2}~[~A_I(\Om)+B_I(\Om) \cos (\phi)+B_R(\Om) \sin 
(\phi) ~],~~~~ \label{osccurr}
\eea
where $A_R,~A_I,~B_R,~B_I$ are the real and imaginary parts of $A$ and
$B$ respectively and $D(\Om)=2 \g^2 \Om^2 k_c/Z(\Om)$ where 
$Z(\Om)= |(k+k_c-m \Om^2- i \g \Om)^2 -k_c^2|^2$.  
From the expressions in Eq.~(\ref{osccurr}) it is clear  that the heat
transfer rates  can 
be separated into deterministic parts (depending on the driving
strength $f_0$) and noisy parts (dependent on temperature of the two
reservoirs). The work terms are temperature independent.  
We now note that the deterministic parts of $\dot{q}_L$ and $\dot{q}_R$,
are both negative.  This can be shown by using the facts that $A_I
\geq 0$ and $  A_I^2-B_I^2-D^2= \g^2\Om^2 [ (k+k_c-m \Om^2)^2+\g^2
  \Om^2 -k_c^2]^2/Z^2 \geq 0 $. This means that for $T_L > T_R$, we
always get $\dot{q}_R < 0$ and hence we can never have heat transfer
from the cold to the hot reservoir. Thus this \emph{cannot}  work as a heat pump.
Also we note that while $\dot{w}_L$ and $\dot{w}_R$ can individually be
negative, the total work done $\dot{w}_L+\dot{w}_R$ is always  
positive. This means that this model \emph{cannot} work as an engine either. These conclusions
remain unchanged even if we define work as $\dot{W}_L = \la f_{L} \dot{x}_1 \ra $,
$\dot{W}_R = \la f_{R} \dot{x}_2 \ra$.

\begin{figure}
\includegraphics[width=3in]{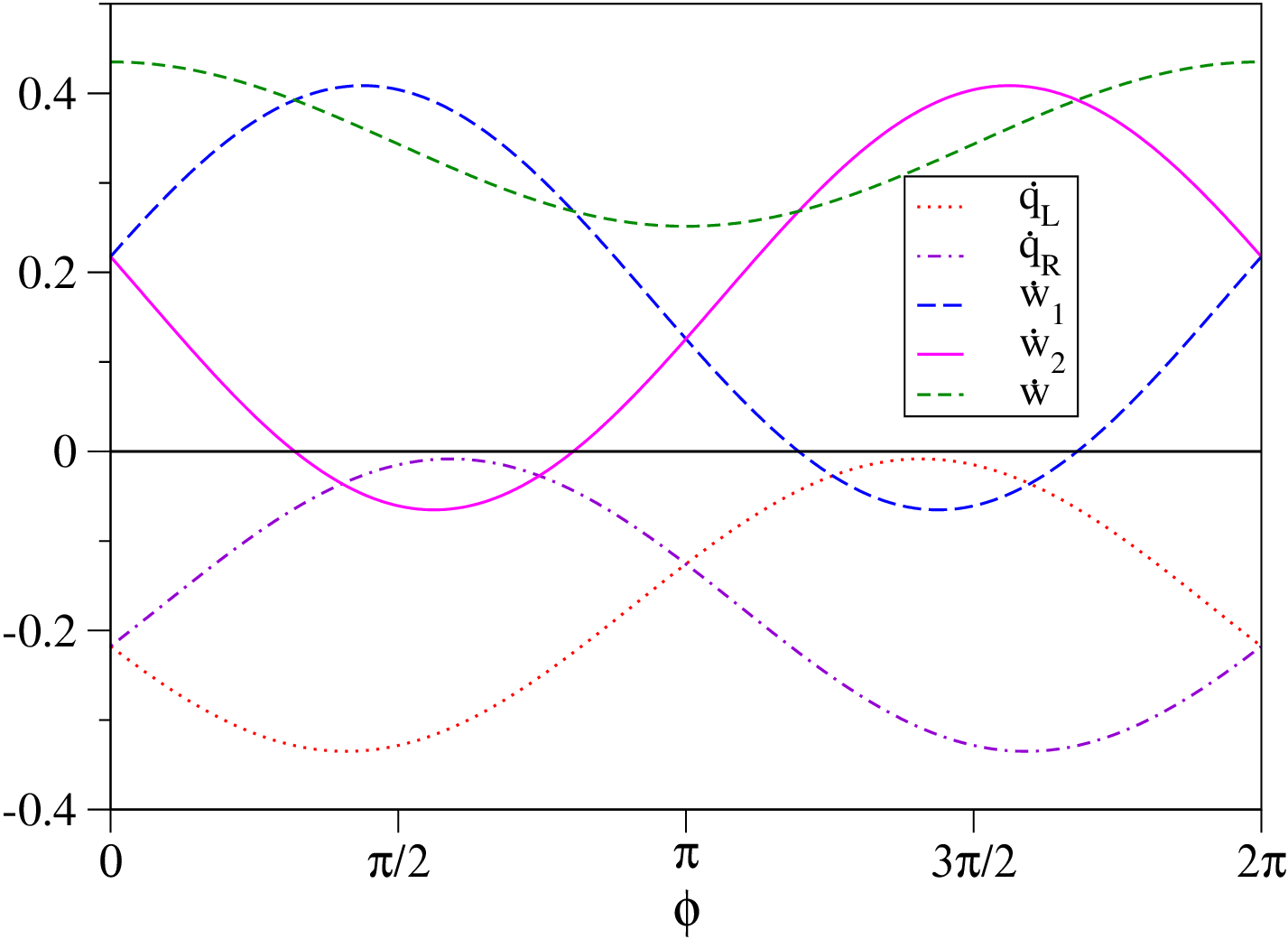}
\caption{(Color online) Plots of heat transfer and work done as a function
  of phase difference $\phi$ in the two particle model. Here $\Om =2\pi/3 $.} 
\label{phi1}
\end{figure} 
In Fig.~(\ref{phi1}) we plot the dependence of the rates
of heat transfer and work done  in the system on the 
phase difference $\phi$. The figures correspond to
the parameter values $k=2,~k_c=3,~m=1,~f_0=1,~\gamma=1$ and
$T_L=T_R=T$. The plots are independent of the temperature $T$. 
Note that the only effect of the driving is to pump in energy which is
asymmetrically distributed between the two reservoirs. 
The asymmetric energy transfer into the baths is an interesting effect
considering that there is no inbuilt directional asymmetry in the system. 

In this model the heat baths and the external driving seem to act
independently on the system.   
It is clear that the linearity of the model leads to this separability
of the effects of the driving and noise forces and this could be the
reason that the model is not able to function as a  heat pump. 
Hence it is important to consider the effect of nonlinearity. We have
numerically studied the effect of including a nonlinear part 
of the form $\alpha [ x_1^4+ x_2^4 +(x_1-x_2)^4 ]/4$ in the oscillator
Hamiltonian. From simulations with a large range of parameter values
we find that the basic conclusions remain unchanged and the model does
not work either as a pump or as an engine.

In conclusion,  we have studied two models which 
have the same ingredients as those on which recent models of quantum
pumps have been constructed.
We find that the first model performs as a heat pump
to transfer heat from a cold to a hot reservoir. Thus pumping is not
an essentially quantum-mechanical phenomena. Also
our model performs as an engine to do  work on the driving
force. It is useful to compare our model with the other well-studied
microscopic model of a engine, namely the Feynman
pawl-and-ratchet. Recent detailed studies have shown that this model
can function 
both as an engine and as a refrigerator \cite{BroeckPRL04}.  
One difference of this
model from ours is that there is no periodic external driving. However
this also means that in order for the model to work in a cyclic way,
at least one of the degrees of freedom has to be a periodic (or
angular) variable.  This may not always be a desirable feature in 
realistic models. 
Surprisingly our second model, though apparently built on the
same  principles, fails to perform either as a pump or as an engine.

The important difference between microscopic models of heat engines,
such as those studied here, and  usual thermodynamic heat engines
is that here the effects of thermal fluctuations are important.
A second difference is that here the system is simultaneously in contact
with both the cold and hot baths. The understanding of these microscopic
models requires the use of nonequilibrium statistical mechanics
and there are currently no general principles as in classical thermodynamics. 
It is clear that further studies are necessary to understand the
pumping mechanism in simple models of molecular pumps and this can
perhaps lead to more realistic and  practical models of molecular
pumps and engines. 

\begin{thebibliography}{**}
\expandafter\ifx\csname natexlab\endcsname\relax\def\natexlab#1{#1}\fi
\expandafter\ifx\csname bibnamefont\endcsname\relax
  \def\bibnamefont#1{#1}\fi
\expandafter\ifx\csname bibfnamefont\endcsname\relax
  \def\bibfnamefont#1{#1}\fi
\expandafter\ifx\csname citenamefont\endcsname\relax
  \def\citenamefont#1{#1}\fi
\expandafter\ifx\csname url\endcsname\relax
  \def\url#1{\texttt{#1}}\fi
\expandafter\ifx\csname urlprefix\endcsname\relax\def\urlprefix{URL }\fi
\providecommand{\bibinfo}[2]{#2}
\providecommand{\eprint}[2][]{\url{#2}}

\bibitem[{\citenamefont{Feynman et~al.}(vol.1, chap. 46,
  1966)\citenamefont{Feynman, Leighton, and Sands}}]{FeynmanLect}
\bibinfo{author}{\bibfnamefont{R.}~\bibnamefont{Feynman}},
  \bibinfo{author}{\bibfnamefont{R.}~\bibnamefont{Leighton}}, \bibnamefont{and}
  \bibinfo{author}{\bibfnamefont{M.}~\bibnamefont{Sands}},
  \emph{\bibinfo{title}{The Feynman Lectures on Physics}}
  (\bibinfo{publisher}{Adison-Wesley}, \bibinfo{address}{Reading
  Massachussets}, \bibinfo{year}{vol.1, chap. 46, 1966}).

\bibitem[{\citenamefont{{Parrondo} and {Espa{\~n}ol}}(1996)}]{Parrondo96}
\bibinfo{author}{\bibfnamefont{J.~M.~R.} \bibnamefont{{Parrondo}}}
  \bibnamefont{and}
  \bibinfo{author}{\bibfnamefont{P.}~\bibnamefont{{Espa{\~n}ol}}},
  \bibinfo{journal}{Am.~J.~Phys.} \textbf{\bibinfo{volume}{64}},
  \bibinfo{pages}{1125} (\bibinfo{year}{1996}); \bibinfo{author}{\bibnamefont{M.O.Magnasco}} \bibnamefont{and}
  \bibinfo{author}{\bibnamefont{G.Stolovitzky}}, \bibinfo{journal}{J. Stat.
  Phys.} \textbf{\bibinfo{volume}{93}}, \bibinfo{pages}{615}
  (\bibinfo{year}{1998}).

\bibitem[{\citenamefont{Sekimoto}(1997)}]{Sekimoto97}
\bibinfo{author}{\bibfnamefont{K.}~\bibnamefont{Sekimoto}},
  \bibinfo{journal}{Jr. Phy. Soc. Jpn.} \textbf{\bibinfo{volume}{66}},
  \bibinfo{pages}{1234} (\bibinfo{year}{1997}); \bibinfo{author}{\bibfnamefont{T.}~\bibnamefont{{Hondou}}} \bibnamefont{and}
  \bibinfo{author}{\bibfnamefont{F.}~\bibnamefont{{Takagi}}},
  \bibinfo{journal}{J. Phy. Soc. Jpn.} \textbf{\bibinfo{volume}{67}},
  \bibinfo{pages}{2974} (\bibinfo{year}{1998}); \bibinfo{author}{\bibfnamefont{H.}~\bibnamefont{{Sakaguchi}}},
  \bibinfo{journal}{J. Phy. Soc. Jpn.} \textbf{\bibinfo{volume}{67}},
  \bibinfo{pages}{709} (\bibinfo{year}{1998}); \bibinfo{author}{\bibfnamefont{C.}~\bibnamefont{{Jarzynski}}} \bibnamefont{and}
  \bibinfo{author}{\bibfnamefont{O.}~\bibnamefont{{Mazonka}}},
  \bibinfo{journal}{\pre} \textbf{\bibinfo{volume}{59}}, \bibinfo{pages}{6448}
  (\bibinfo{year}{1999});
Y. Lee \emph{et al.}, Phys. Rev. Lett. {\bf 91}, 220601 (2003).

\bibitem[{\citenamefont{Magnasco}(1993)}]{Magnasco93}
\bibinfo{author}{\bibfnamefont{M.~O.} \bibnamefont{Magnasco}},
  \bibinfo{journal}{Phys. Rev. Lett.} \textbf{\bibinfo{volume}{71}},
  \bibinfo{pages}{1477} (\bibinfo{year}{1993}); 
I. Der\'enyi and T. Vicsek, Phys. Rev. Lett. {\bf 75}, 374 (1995);
\bibinfo{author}{\bibfnamefont{F.}~\bibnamefont{J\"ulicher}},
  \bibinfo{author}{\bibfnamefont{A.}~\bibnamefont{Ajdari}}, \bibnamefont{and}
  \bibinfo{author}{\bibfnamefont{J.}~\bibnamefont{Prost}},
  \bibinfo{journal}{Rev. Mod. Phys.} \textbf{\bibinfo{volume}{69}},
  \bibinfo{pages}{1269} (\bibinfo{year}{1997}); \bibinfo{author}{\bibfnamefont{P.}~\bibnamefont{Reimann}},
  \bibinfo{journal}{Physics Reports} \textbf{\bibinfo{volume}{361}},
  \bibinfo{pages}{57} (\bibinfo{year}{2002}); \bibinfo{author}{\bibfnamefont{R.~D.} \bibnamefont{Astumian}},
  \bibinfo{journal}{Science} \textbf{\bibinfo{volume}{276}},
  \bibinfo{pages}{917 } (\bibinfo{year}{1997}).

\bibitem[{\citenamefont{Feldmann et~al.}(1996)\citenamefont{Feldmann, Geva,
  Kosloff, and Salamon}}]{FeldmannHeatEngine96}
\bibinfo{author}{\bibfnamefont{T.}~\bibnamefont{Feldmann}},
  \bibinfo{author}{\bibfnamefont{E.}~\bibnamefont{Geva}},
  \bibinfo{author}{\bibfnamefont{R.}~\bibnamefont{Kosloff}}, \bibnamefont{and}
  \bibinfo{author}{\bibfnamefont{P.}~\bibnamefont{Salamon}},
  \bibinfo{journal}{Am.~J.~Phys.} \textbf{\bibinfo{volume}{64}},
  \bibinfo{pages}{485} (\bibinfo{year}{1996});
\bibinfo{author}{\bibfnamefont{I.}~\bibnamefont{{Der{\'e}nyi}}}
  \bibnamefont{and} \bibinfo{author}{\bibfnamefont{R.~D.}
  \bibnamefont{{Astumian}}}, \bibinfo{journal}{\pre}
  \textbf{\bibinfo{volume}{59}}, \bibinfo{pages}{6219} (\bibinfo{year}{1999}); \bibinfo{author}{\bibfnamefont{A.}~
  \bibnamefont{{Gomez-Marin}}}
  \bibnamefont{and} \bibinfo{author}{\bibfnamefont{J.~M.}
  \bibnamefont{{Sancho}}}, \bibinfo{journal}{\pre}
  \textbf{\bibinfo{volume}{71}}, \bibinfo{pages}{021101}
  (\bibinfo{year}{2005}); 
\bibinfo{author}{\bibfnamefont{M.~C.} \bibnamefont{{Mahato}}},
  \bibinfo{author}{\bibfnamefont{T.~P.} \bibnamefont{{Pareek}}},
  \bibnamefont{and} \bibinfo{author}{\bibfnamefont{A.~M.}
  \bibnamefont{{Jayannavar}}}, \bibinfo{journal}{Int. J.
  Mod. Phys. B} \textbf{\bibinfo{volume}{10}}, \bibinfo{pages}{3857}
  (\bibinfo{year}{1996}). 



\bibitem[{\citenamefont{Astumian and Derényi}(1997)}]{Astumian97}
\bibinfo{author}{\bibfnamefont{R.~D.} \bibnamefont{Astumian}} \bibnamefont{and}
  \bibinfo{author}{\bibfnamefont{I.}~\bibnamefont{Derényi}},
  \bibinfo{journal}{European Biophysics Journal} \textbf{\bibinfo{volume}{27}},
  \bibinfo{pages}{474} (\bibinfo{year}{1997}).

\bibitem[{\citenamefont{{Buttiker} et~al.}(1994)\citenamefont{{Buttiker},
  {Thomas}, and {Pretre}}}]{Buttiker94}
\bibinfo{author}{\bibfnamefont{M.}~\bibnamefont{{Buttiker}}},
  \bibinfo{author}{\bibfnamefont{H.}~\bibnamefont{{Thomas}}}, \bibnamefont{and}
  \bibinfo{author}{\bibfnamefont{A.}~\bibnamefont{{Pretre}}},
  \bibinfo{journal}{Z. Phys. B.} \textbf{\bibinfo{volume}{94}},
  \bibinfo{pages}{133} (\bibinfo{year}{1994}); \bibinfo{author}{\bibfnamefont{P.~W.} \bibnamefont{{Brouwer}}},
  \bibinfo{journal}{\prb} \textbf{\bibinfo{volume}{58}}, \bibinfo{pages}{10135}
  (\bibinfo{year}{1998}); \bibinfo{author}{\bibfnamefont{L.}~\bibnamefont{{Arrachea}}},
  \bibinfo{journal}{\prb} \textbf{\bibinfo{volume}{72}},
  \bibinfo{pages}{125349} (\bibinfo{year}{2005}{\natexlab{a}}); \bibinfo{author}{\bibfnamefont{L.}~\bibnamefont{{Arrachea}}},
  \bibinfo{journal}{\prb} \textbf{\bibinfo{volume}{72}},
  \bibinfo{pages}{121306(R)} (\bibinfo{year}{2005}{\natexlab{b}}); \bibinfo{author}{\bibfnamefont{M.}~\bibnamefont{{Strass}}},
  \bibinfo{author}{\bibfnamefont{P.}~\bibnamefont{{H{\"a}nggi}}},
  \bibnamefont{and} \bibinfo{author}{\bibfnamefont{S.}~\bibnamefont{{Kohler}}},
  \bibinfo{journal}{\prl} \textbf{\bibinfo{volume}{95}},
  \bibinfo{pages}{130601} (\bibinfo{year}{2005}); \bibinfo{author}{\bibfnamefont{S.}~\bibnamefont{{Kohler}}},
  \bibinfo{author}{\bibfnamefont{J.}~\bibnamefont{{Lehmann}}},
  \bibnamefont{and}
  \bibinfo{author}{\bibfnamefont{P.}~\bibnamefont{{H{\"a}nggi}}},
  \bibinfo{journal}{Phy. Rep.} \textbf{\bibinfo{volume}{406}},
  \bibinfo{pages}{379} (\bibinfo{year}{2005}); \bibinfo{author}{\bibfnamefont{A.}~\bibnamefont{Agarwal}} \bibnamefont{and}
  \bibinfo{author}{\bibfnamefont{D.}~\bibnamefont{Sen}}, 
  \bibinfo{journal}{J. Phys. Condens. Matter} \textbf{\bibinfo{volume}{19}},
  \bibinfo{pages}{046205} (\bibinfo{year}{2007}).

\bibitem[{\citenamefont{{Switkes}}(1999)}]{SwitkesSciqpumpexp99}
\bibinfo{author}{\bibfnamefont{M.}~\bibnamefont{{Switkes}}} 
  \bibinfo{author}{\emph{et ~al.}}, \bibinfo{journal}{Science}
  \textbf{\bibinfo{volume}{283}}, \bibinfo{pages}{1905} (\bibinfo{year}{1999}); \bibinfo{author}{\bibfnamefont{P.}~\bibnamefont{Leek}} 
 \bibinfo{author}{\emph{et ~al.}}, \bibinfo{journal}{\prl}
  \textbf{\bibinfo{volume}{95}}, \bibinfo{pages}{256802}
  (\bibinfo{year}{2005}).

\bibitem[{\citenamefont{{Segal} and {Nitzan}}(2006)}]{Nitzan06}
\bibinfo{author}{\bibfnamefont{D.}~\bibnamefont{{Segal}}} \bibnamefont{and}
  \bibinfo{author}{\bibfnamefont{A.}~\bibnamefont{{Nitzan}}},
  \bibinfo{journal}{\pre} \textbf{\bibinfo{volume}{73}},
  \bibinfo{pages}{026109} (\bibinfo{year}{2006}).

\bibitem[{\citenamefont{{Glauber}}(1963)}]{Glauber63}
\bibinfo{author}{\bibfnamefont{R.}~\bibnamefont{{Glauber}}},
  \bibinfo{journal}{Jr. Math. Phys.} \textbf{\bibinfo{volume}{4}},
  \bibinfo{pages}{294} (\bibinfo{year}{1963}).

\bibitem[{\citenamefont{{van den Broeck} et~al.}(2004)\citenamefont{{van den
  Broeck}, {Kawai}, and {Meurs}}}]{BroeckPRL04}
\bibinfo{author}{\bibfnamefont{C.}~\bibnamefont{{van den Broeck}}},
  \bibinfo{author}{\bibfnamefont{R.}~\bibnamefont{{Kawai}}}, \bibnamefont{and}
  \bibinfo{author}{\bibfnamefont{P.}~\bibnamefont{{Meurs}}},
  \bibinfo{journal}{\prl} \textbf{\bibinfo{volume}{93}},
  \bibinfo{pages}{090601} (\bibinfo{year}{2004}); \bibinfo{author}{\bibfnamefont{C.}~\bibnamefont{Van~den Broeck}}
  \bibnamefont{and} \bibinfo{author}{\bibfnamefont{R.}~\bibnamefont{Kawai}},
  \bibinfo{journal}{Phys. Rev. Lett.} \textbf{\bibinfo{volume}{96}},
  \bibinfo{pages}{210601} (\bibinfo{year}{2006}).



\end{thebibliography}

\end{document}